\documentclass[pdflatex,sn-mathphys-num]{sn-jnl}
\usepackage[utf8]{inputenc}
\usepackage[T1]{fontenc}
\usepackage{lmodern}
\usepackage{graphicx}
\usepackage{multirow}
\usepackage{booktabs}
\usepackage{longtable}
\usepackage{tabularx}
\usepackage{array}
\usepackage{adjustbox}
\usepackage{threeparttable}
\usepackage{amsmath,amssymb,amsfonts}
\usepackage{amsthm}
\usepackage{mathrsfs}
\usepackage{siunitx}
\usepackage{xcolor}
\usepackage{textcomp}
\usepackage{ragged2e}
\usepackage{geometry}
\usepackage{float}
\usepackage{placeins}
\usepackage{lineno}
\usepackage{algorithm}
\usepackage{algorithmicx}
\usepackage{algpseudocode}
\usepackage{listings}
\usepackage[title]{appendix}
\usepackage{manyfoot}
\usepackage{pdflscape}
\usepackage[hypcap=false]{caption}
\usepackage{hyperref}
\theoremstyle{thmstyleone}

\theoremstyle{thmstyletwo}

\theoremstyle{thmstylethree}

\DeclareUnicodeCharacter{2212}{-}
\begin{document}

\title{Deep Learning-Driven Downscaling for Climate Risk Assessment of Projected Temperature Extremes in the Nordic Region}

\author*[1]{\fnm{Parthiban} \sur{Loganathan}}\email{parthi@kth.se}
\author[1]{\fnm{Elias} \sur{Zea}}
\author[2]{\fnm{Ricardo} \sur{Vinuesa}}
\author[1]{\fnm{Evelyn} \sur{Otero}}

\affil[1]{\orgdiv{Department of Engineering Mechanics}, \orgname{KTH Royal Institute of Technology}, \orgaddress{\city{Stockholm}, \country{Sweden}}}
\affil[2]{\orgdiv{Department of Aerospace Engineering}, \orgname{University of Michigan}, \orgaddress{\city{Ann Arbor}, \country{United States}}}

\abstract{

Rapid changes and increasing climatic variability across the widely varied Köppen-Geiger regions of northern Europe generate significant needs for adaptation. Regional planning needs high-resolution projected temperatures. This work presents an integrative downscaling framework that incorporates Vision Transformer (ViT), Convolutional Long Short-Term Memory (ConvLSTM), and Geospatial Spatiotemporal Transformer with Attention and Imbalance-Aware Network (GeoStaNet) models. The framework is evaluated with a multicriteria decision system, Deep Learning-TOPSIS (DL-TOPSIS), for ten strategically chosen meteorological stations encompassing the temperate oceanic (Cfb), subpolar oceanic (Cfc), warm-summer continental (Dfb), and subarctic (Dfc) climate regions. Norwegian Earth System Model (NorESM2-LM) Coupled Model Intercomparison Project Phase 6 (CMIP6) outputs were bias-corrected during the 1951-2014 period and subsequently validated against earlier observations of day-to-day temperature metrics and diurnal range statistics. The ViT showed improved performance (Root Mean Squared Error (RMSE): $1.01\,^\circ$C; $R^2$: 0.92), allowing for production of credible downscaled projections. Under the SSP5-8.5 scenario, the Dfc and Dfb climate zones are projected to warm by $4.8\,^\circ$C and $3.9\,^\circ$C, respectively, by 2100, with expansion in the diurnal temperature range by more than $1.5\,^\circ$C. The Time of Emergence signal first appears in subarctic winter seasons (Dfc: $\sim$2032), signifying an urgent need for adaptation measures. The presented framework offers station-based, high-resolution estimates of uncertainties and extremes, with direct uses for adaptation policy over high-latitude regions with fast environmental change.}


\maketitle

\section{Introduction}

The Nordic countries, comprising Denmark, Finland, Iceland, Norway, and Sweden, are a natural experimental laboratory for monitoring and quantifying the profound behaviour of anthropogenic climate change. The region is not only critical for tracking large-scale climate signals but also for understanding the atmospheric conditions that govern processes like aircraft contrail formation. The extensive latitudinal gradient spanning from temperate oceanic conditions in southern Denmark to subarctic conditions in northern Scandinavia serves as an optimal field for assessing climate system behaviour in response to the influence of humans. This large area includes four different Köppen-Geiger classes: a) Cfb, b) Cfc, c) Dfb, and d) Dfc \cite{Beck2023, Stefansson2024}. Greater insight into climatic conditions across such a range of heterogeneous conditions provides important regional climate sensitivities information and important background for adaptation planning for the entire circumpolar north.

Recent scientific studies have consistently shown that warming in the Nordic region is significantly higher than the global average, primarily due to Arctic amplification processes. These processes are made more difficult by feedback systems, including reduced sea-ice coverage, alterations in surface albedo, modifications in atmospheric and oceanic circulation patterns, and variations in energy balance dynamics \cite{Rantanen2022, Pistone2019}. As a result, broad-scale meteorological networks display systematic increases in annual mean temperatures, a rise in the occurrence and magnitude of heat waves, lengthened growing seasons, and considerable shifts in precipitation regimes across all Nordic countries \cite{Donat2013}. The main concerns are clear changes in temperature extremes, such as soaring summer highs, changed winter lows, and expanding daily temperature ranges. These shifts in temperature extremes are especially important because they directly affect atmospheric conditions that determine the formation and persistence of aircraft contrails. All of these things point to big changes in the climate of the region. These complicated changes show how important it is to know how the climate changes in different places.

 The complex spatial nature of climatic changes in the Nordic region poses scientific obstacles and possibilities for enhanced comprehension of climate mechanisms. Patterns of warming along the coast are different from patterns of warming in the continental interior. Elevation changes make it harder to predict temperature and precipitation patterns \cite{ECMWF2024}. There are a lot of different climate responses because of the mountains, long coastlines, numerous islands, and vast inland waters. This means that spatially explicit analytical methods are needed. Such detailed approaches are also essential for capturing localized atmospheric conditions that are important for processes like contrail occurrence, which depend on precise spatial variations in temperature and humidity. This is particularly relevant for aviation operations, as contrail formation and persistence can vary sharply over short distances in response to subtle climatic gradients. New developments in high-resolution gridded climate datasets, coupled with standardisation of monitoring networks, have ensured that researchers are much better at discerning spatial deviations and temporal trends \cite{Klein2009, ECAD2025}. Coupling high-resolution observations with advanced modelling techniques allows us, for the first time, to elucidate past climate variability and climate changes in the future with high accuracy across the Nordic region.

The development of regional climate scenarios has undergone a substantial transformation, driven by improvements in computational power and the development of methodologies. Current approaches demonstrate the need for a standardised model selection procedure for General Circulation Models (GCMs), grounded in comprehensive performance evaluations rather than capricious model choices \cite{IPCC2021, IPCC2022}. The design guidelines for station networks now focus on achieving optimal spatial representation across the most critical climate regions, while advanced clustering and classification methodologies facilitate the reduction of redundancy \cite{Peel2007, Stefansson2024}. This advancement has been supported by widespread global data-sharing efforts, most notably the European Climate Assessment and Dataset (ECAD), which provides the E-OBS daily gridded meteorological dataset for Europe, available via the Copernicus Climate Data Store (CDS) \cite{ECAD2025}. The E-OBS collection provides quality-controlled, standardised gridded temperature and precipitation data essential for rigorous climate research.

The development of GCM evaluation methodologies has revolutionised regional climate model fields by encouraging objective, multi-metric assessment platforms. Systematic studies of model performance, including various climate variables, seasonal variation, and spatial resolutions, are now available in performance maps and for comprehensive benchmark studies. Among the CMIP6 model collection, the Norwegian Earth System Model version 2 (NorESM2-LM) has repeatedly demonstrated exceptional ability for replicating Nordic climate conditions; it stands out most notably for its accuracy in temperature predictions and reproduction of circulation patterns, and depiction of extremities of climate conditions \cite{Loganathan2025, Loganathan2021b}. Advanced model selection procedures based on Multi-Criteria Decision-Making (MCDM) tools, such as Deep-Learning-TOPSIS (DL-TOPSIS), offer sound, reproducible techniques for determining best-performing models with minimum subjective bias \cite{Talukder2025, Brunner2020}.

Despite computational efficiency, customary statistical downscaling methods often fail to adequately capture the complex, nonlinear interactions between large-scale atmospheric conditions and local climatic outcomes. This limitation is especially apparent in regions like the Nordics, where intricate geography and pronounced maritime-continental gradients present unique modelling challenges. Traditional techniques-including linear regression, quantile mapping, and delta change methods-are frequently prone to systematic errors in representing extremes and may lose temporal and spatial coherence \cite{Teutschbein2018}.

In response to these challenges, recent advances in machine learning and deep learning have transformed the capabilities of downscaling. Ensemble approaches such as random forests and gradient boosting models (e.g., XGBoost), neural network methods such as multilayer perceptrons, and sophisticated deep architectures like convolutional neural networks, LSTM models, ViT, and the climate-specific GeoStaNet now set new standards for accuracy and reliability \cite{Breiman2001, Reichstein2019, Jardines2024}. These techniques create new opportunities for high-resolution climate projections essential for regional adaptation. 

 Performance evaluation protocols have also advanced, now including comprehensive multi-metric evaluation frameworks that quantify model competency across multiple facets of climate faithfulness. State-of-the-art evaluation methodologies combine standard error metrics with distribution accuracy measures, temporal consistency metrics, extremal depiction capabilities, and climate signal detection metrics \cite{Chai2014, Hawkins2009}. Entropy-weighted multi-criteria aggregation procedures provide a fair evaluation across diversified performance axes while incorporating relevant climate impacts' indices, like heat wave count, frost day frequency, seasonal mean temperature variation, and time-of-emergence metrics \cite{Loganathan2025, Zhang2011, Klein2009, Alexander2006}. Such comprehensive evaluation frameworks form a scientifically viable basis for decision-making, therefore directly contributing to climate risk assessment, adaptation planning, and policy development \cite{Loganathan2021b}.
 
 Integrating sophisticated downscaling technologies with comprehensive evaluation frameworks and best-performing GCM selection procedures offers unparalleled opportunities for high-resolution, policy-relevant climate scenario product delivery. This integration of methodologies permits high-resolution climate projections that are both spatially and temporally comprehensive. Such climate projections may inform adaptation courses of action across various sectors-agriculture, forestry, water resources, infrastructure, power production, and public health-across the Nordic countries \cite{Loganathan2020a, Loganathan2021a}. State-of-the-art climate services now require scenario products that are not just scientifically viable but also practically useful towards decision-making, demanding facile integration of climate science with practical application.

 The present work follows an integrated methodology structure by developing and applying a complete framework that synthesises optimal station network designs, strict selections of GCMs, deep-learning downscaled outputs, and multi-criteria performance evaluations. A carefully chosen ten-station meteorological network has been strategically located to encompass all major Nordic climatic regimes, with resultant representative coverage of regional climatic variability. NorESM2-LM acts as the key GCM input, selected based on its previously established superior performance in thorough evaluations\cite{Loganathan2025}. Numerous methodologies of downscaled outputs, such as classical statistics-based approaches, machine learning-based models, and advanced deep-learning architectures, are exhaustively analysed under DL-TOPSIS multi-criteria ranking methodologies. Climate projections for a moderate SSP2-4.5 and a high SSP5-8.5 emission scenario are generated, with time divisions covering the near term (2015-2050), the mid-century (2051-2075), and the end of the century (2076-2100). This systematic exercise allows the production of rigorously scientific and spatially explicit and temporally precise climate scenario datasets that may directly inform initiatives related to climate adaptation and climate resilience planning for the Nordic region.

\section{Study Area and Data Description}
\begin{figure}[!htbp]
 \centering
 \includegraphics[width=1\textwidth]{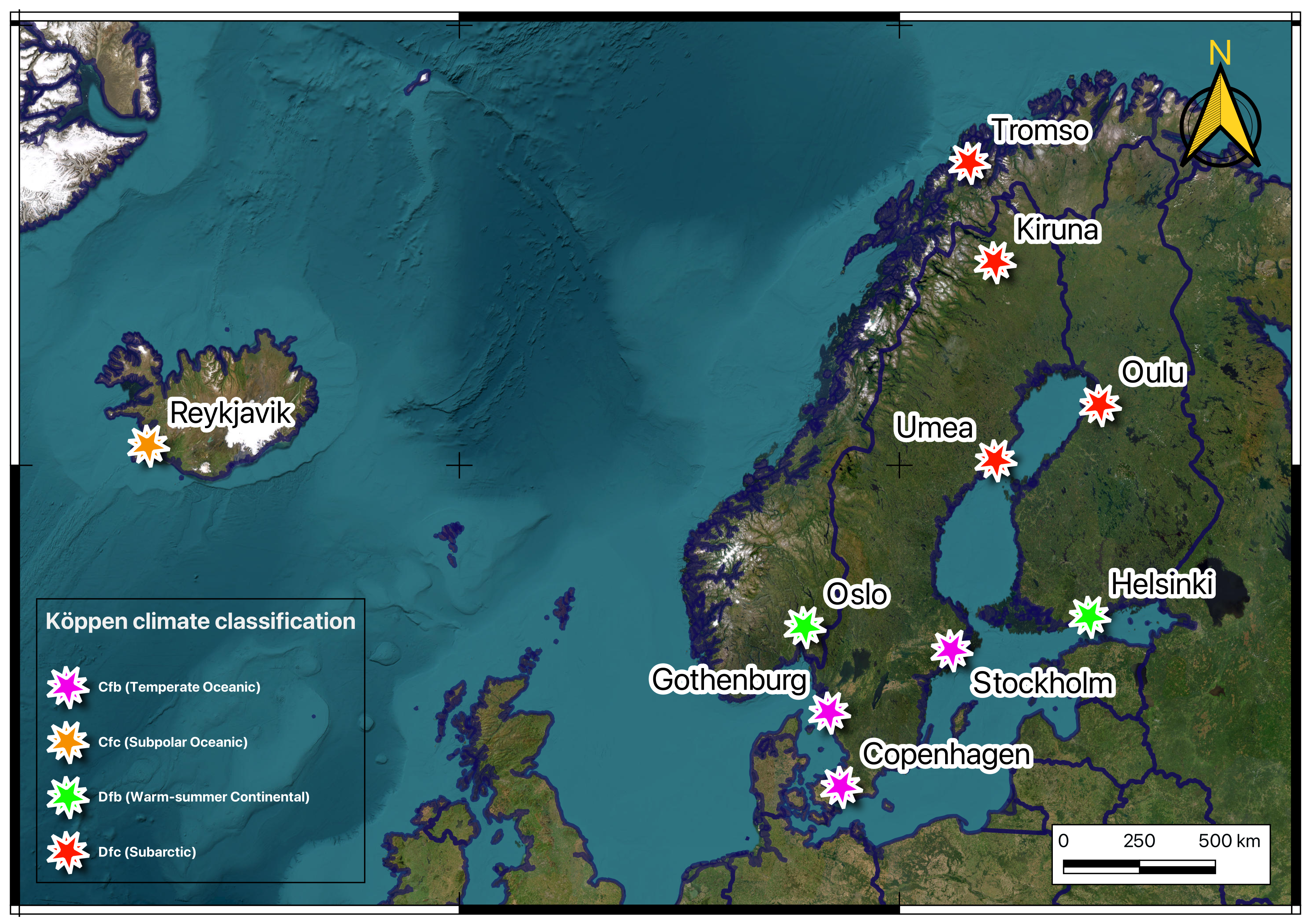}
 \caption{Geographical distribution of the ten selected Nordic stations, colour-coded by Köppen-Geiger climate classification.}
 \label{fig:1}
\end{figure}

The Nordic region has a remarkable range of climatic conditions, mediated by its great latitudinal and longitudinal dispersion, complicated topography, and its proximity to the Atlantic and Arctic Oceans. This paper focuses on ten selected meteorological stations-Copenhagen, Stockholm, Gothenburg, Reykjavik, Helsinki, Oslo, Kiruna, Oulu, Tromsø, and Umeå. They were selected consciously to exhibit the four major Köppen-Geiger climate categories: Cfb, Cfc, Dfb, and Dfc (Fig.~\ref{fig:1}) \cite{Beck2023}. As illustrated in Table~\ref {tab:1}, the chosen stations span latitudes between 55.68°N and 69.65°N and longitudes between 12.57°E and 21.94°W and range from 9 to 452 metres in altitude. This selection successfully covers a whole range of gradients, from maritime over continental to lowland to subarctic alpine-influenced climates.

The choice of stations was governed by a multivariate clustering of extensive climatological means with regard to rainfall and temperature. This method ensures minimal redundancy with optimal representation of diverse macroclimates \cite{Peel2007, Stefansson2024}. This involved performing hierarchical agglomerative clustering on standardised anomalies with follow-up silhouette analysis for verification of cluster stability. This ultimate network, in addition to encompassing key transition areas, for example, that between continental and temperate climates across central Sweden and central Finland, also encompasses harsh subarctic regions of northern Scandinavia, which are useful for the detection of extreme events as much as for ascertaining thresholds of emergence. 

Daily maximum temperature (\textrm{tasmax}) and minimum temperature (\textrm{tasmin}) observational data for the 1951-2014 period were sourced from national meteorological agencies and from the European Climate Assessment Dataset (ECAD) \cite{ECAD2025}. Quality control overlaid automated outlier detection, seasonal autoregressive estimation for missing values, and a homogenisation method for correcting for station relocations and changes in measuring instruments \cite{Wilby2004, ECMWF2024}. This dataset was aligned with NorESM2-LM history outputs for the period 1951-2014 for deriving common baselines for training and validating the downscaling models \cite{Seland2020, Loganathan2025}.

Future climate outputs for Shared Socioeconomic Pathways SSP2-4.5 and SSP5-8.5 were computed from NorESM2-LM CMIP6 simulations for the 2015-2100 period. Such projections were classified as near-term (2015-2050), mid-term (2051-2075), and far-future (2076-2100) to elucidate sequential climatic responses \cite{VanVuuren2011, Tebaldi2021}. Underlying variables include both the daily \textrm{tasmax}, \textrm{tasmin}, and the diurnally computed diurnal temperature range (\textrm{dtr} = \textrm{tasmax} - \textrm{tasmin}), allowing a thorough analysis of mean conditions, extremities, and variability. 

\begin{table}[!htbp]
 \centering
 \caption{Details of the ten Nordic stations used in this study.}
 \label{tab:1}
 \begin{tabular}{llrrrl}
 \hline
 \textbf{Station} & \textbf{Country} & \textbf{Lat (°N)} & \textbf{Lon (°E)} & \textbf{Zone} & \textbf{Elev (m)} \\
 \hline
 Copenhagen & Denmark & 55.68 & 12.57 & Cfb & 9 \\
 Gothenburg & Sweden & 57.71 & 11.97 & Cfb & 12 \\
 Stockholm & Sweden & 59.33 & 18.07 & Cfb & 44 \\
 Reykjavik & Iceland & 64.15 & -21.94 & Cfc & 61 \\
 Helsinki & Finland & 60.17 & 24.94 & Dfb & 25 \\
 Oslo & Norway & 59.91 & 10.75 & Dfb 
& 94 \\
 Kiruna & Sweden & 67.86 & 20.23 & Dfc & 452 \\
 Oulu & Finland & 65.01 & 25.47 & Dfc & 15 \\
 Tromsø & Norway & 69.65 & 18.96 & Dfc & 10 \\
 Umeå & Sweden & 63.83 & 20.26 & Dfc & 10 \\
 \hline
 \end{tabular}
\end{table}

The station network, supported by stringent selection and quality assurance processes, guarantees a thorough representation of the Nordic region's climatic diversity. This dataset constitutes the essential foundation for subsequent modelling, assessment, and multi‑criteria ranking across both historical and future climate scenarios.

\section{Methodology}
\begin{figure}[!htbp]
 \centering
 \includegraphics[width=1\textwidth]{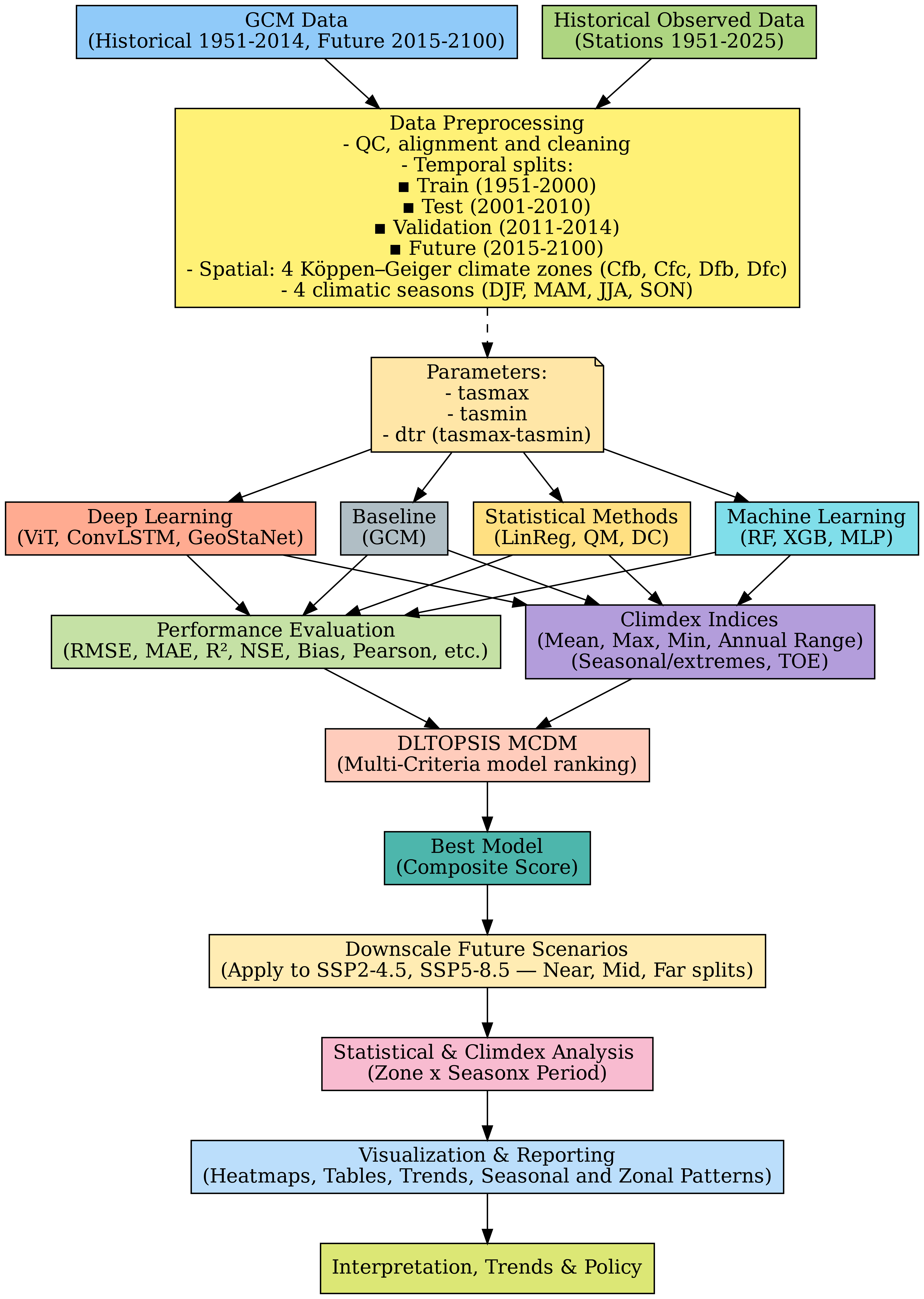}
 \caption{Workflow of the multi‑stage DL-TOPSIS downscaling framework, from GCM and observations through model ranking to scenario projection.}
 \label{fig:2}
\end{figure}

The research utilises a comprehensive, multilevel methodological framework (Fig.~\ref{fig:2}). This framework includes data preprocessing, formulation of models for downscaling across different paradigms, multi-criteria metrics for performance appraisal, and a non-subjective ranking methodology applied via a hybrid Deep-Learning-TOPSIS (DL-TOPSIS) technique. Such a framework ensures methodological correctness and scalability, and interpretability across various climate regimes and time contexts.

\subsection{Data Sources and Preprocessing}
We utilised historical daily maxima and minima temperatures (\textrm{tasmax}, \textrm{tasmin}) from ten Nordic synoptic meteorological stations (1951-2014), sourced from national archives and homogenised with agreed protocols \cite{Wilby2004} and \cite{ECMWF2024}. Quality control measures entailed outlier identification by applying interquartile range filtering, missing value imputation with seasonally stratified autoregressive interpolation, and homogenisation correcting for station shifts and changes in metadata. This observational record was matched with NorESM2-LM CMIP6 outputs \cite{Seland2020, Loganathan2025} for the same historical period (1951-2014) to preserve spatial and temporal consistency. We specified time segments as training period (1951-2000), test period (2001-2010), validation period (2011-2014), and projection period (2015-2100). Spatial classification followed the Köppen-Geiger classification, which includes four zones (Cfb, Cfc, Dfb, Dfc) and four climatic seasons (DJF, MAM, JJA, SON), and \textrm{dtr} is computed. 

\subsection{Downscaling Model Framework}
Ten various downscaling models were constructed, including statistical, machine learning, and deep-learning models, in addition to a GCM benchmark Table~\ref{tab:2}. For equitable comparability of method, the same set of predictors was employed for training each model, with standardised metrics for evaluation.

\begin{landscape}
\begin{table}[p]
\centering
\small
\caption{Downscaling models: formulation, core architecture, and key hyperparameters.}
\label{tab:2}
\setlength{\tabcolsep}{4pt}
\begin{tabular}{@{} p{4.5cm} p{6.5cm} p{6cm} p{6cm} @{}}
\toprule
\textbf{Model} & \textbf{Method / Equations (concise)} & \textbf{Architecture} & \textbf{Key hyperparameters} \\
\midrule

\textbf{Linear Regression} &
$\hat{y}_t = \beta_0 + \sum_{i=1}^p \beta_i x_{i,t} + \varepsilon_t$. Ordinary least squares/ridge. &
Single linear mapping from predictors $\{x_{i,t}\}$ to station value. &
Ridge $\lambda=0.05$, predictors standardised. \cite{Benestad2008} \\[6pt]

\midrule
\textbf{Quantile Mapping (QM)} &
Distributional transfer: $y^*_t = F^{-1}_{obs}\big(F_{mod}(x_t)\big)$ applied monthly. &
Non‑parametric empirical Cumulative Distribution Function (CDF) method (monthly bins). &
500 quantiles/month; seasonal grouping. \cite{Cannon2015} \\[6pt]

\midrule
\textbf{Delta Change (DC)} &
Monthly additive deltas: $y^*_t = y_{obs,t} + \Delta_m$, $\Delta_m=\bar{y}^{fut}_m-\bar{y}^{hist}_m$. &
Applies climatological shift to observations. &
Simple baseline; no variance correction. \cite{Hay2000} \\[6pt]

\midrule
\textbf{Random Forest (RF)} &
Ensemble average of regression trees: $\hat{y}=\frac{1}{N}\sum_k T_k(x)$. &
Scikit‑learn RF regression. &
100 trees, max\_depth=10, min\_samples\_leaf=2. \cite{Breiman2001} \\[6pt]

\midrule
\textbf{XGBoost} &
Gradient boosting on decision trees, minimising stagewise residuals. &
XGBoost regressor with regularisation. &
100 estimators, $\eta=0.1$, max\_depth=8, subsample=0.8. \cite{Chen2016} \\[6pt]

\midrule
\textbf{Multi-Layer Perceptron (MLP) (feedforward)} &
Layered dense mapping: $h^{(l)}=\sigma(W^{(l)}h^{(l-1)}+b^{(l)})$. &
Dense network (fully connected). &
Layers [128, 64, 32], Rectified Linear Unit (ReLU), dropout=0.3, Adam $lr=10^{-4}$.\\[6pt]

\midrule
\textbf{ConvLSTM} &
ConvLSTM gating equations (spatio‑temporal recurrence), e.g., gates use convolution $*$:
$i_t=\sigma(W_{xi}*X_t + W_{hi}*H_{t-1}+b_i)$ etc. &
Stacked ConvLSTM layers processing spatio‑temporal cubes. &
2 layers (filters [32, 64]), kernel=3×3; input window $T=30$; Adam $lr=10^{-4}$. \\[6pt]

\midrule
\textbf{Vision Transformer (ViT)} &
Patch embedding $z_i = E x_i + E_{pos}$; multi‑head self‑attention:
$\mathrm{Attention}(Q,K,V)=\mathrm{softmax}(\frac{QK^\top}{\sqrt{d_k}})V$. &
Patch‑based transformer encoder applied to grid patches. &
Patch=4×4, embed dim=256, 8 encoder blocks, AdamW $lr=3\cdot10^{-5}$. \\[6pt]

\midrule
\textbf{GeoStaNet} &
ViT + geospatial embedding $e_g = W_g[lat,lon]^\top$ and temporal transformer:
final loss $\mathcal{L}=\alpha\mathrm{MSE}_{all}+(1-\alpha)\mathrm{MSE}_{ext}$. &
Hybrid Convolutional Neural Network (CNN)/Transformer: patch embedding + geolocation proj + temporal transformer + upsampling. &
Patch 8×8, temporal encoders=4, $\alpha=0.8$, upsampling via transposed conv. \cite{BanoMedina2022} \\[6pt]

\midrule
\textbf{GCM Baseline} &
Raw NorESM2-LM output interpolated to station locations; no learning. &
Reference baseline. &
Intercomparison baseline: used to quantify downscaling gains. \cite{Seland2020} \\

\bottomrule
\end{tabular}
\end{table}
\end{landscape}

\subsection{Performance Metrics and Evaluation}

To quantify model faithfulness, we utilised ten mutually complementary measures that examine magnitude, variability, and temporally coherent behaviour: Root Mean Squared Error (RMSE), Mean Absolute Error (MAE) \cite{Chai2014}, bias, coefficient of determination ($R^2$) \cite{Gupta2009}, Nash-Sutcliffe Efficiency (NSE) \cite{Nash1970}, Pearson correlation \cite{Loganathan2025}, sign correlation, maximum/minimum relative error, and Time of Emergence (TOE) \cite{Hawkins2012}. All of these measures correspond to established frameworks for measuring skill \cite{Perkins2007}.

\subsection{Hybrid DL-TOPSIS Model Ranking}
A hybrid framework, which integrates TOPSIS with deep learning \cite{Loganathan2025, Talukder2025, Li2024}, allows for unbiased integration of performance measures that are mutually correlated with the application of dynamically changing neural weighting. 

Let \( r_{ij} \) be the normalised performance of the model \( i \) under criterion \( j \), \( v_{ij} \) the weighted normalised score, and \( CC_i \) the closeness coefficient for model \( i \), quantifying its closeness to the ideal solution.

\[
r_{ij} = \frac{x_{ij}}{\sqrt{\sum_i x_{ij}^2}}, \quad
v_{ij} = w_j r_{ij}, \quad
CC_i = \frac{S_i^-}{S_i^+ + S_i^-}
\]

Here, \( x_{ij} \) is the raw performance value of the model \( i \) with respect to criterion \( j \), \( w_j \) is the adaptively learnt weight for criterion \( j \), \( S_i^+ \) and \( S_i^- \) denotes the distances from the ideal and anti-ideal solutions for model \( i \), respectively, and \( CC_i \) is the resulting closeness coefficient.

\subsection{Model Implementation and Hyperparameter Tuning}
Hyperparameter tuning was performed with TensorFlow and PyTorch implementations, which were run on a multi-GPU NVIDIA cluster. Hyperparameter tuning followed a two-stage regime: starting with an initial coarse grid search that was refined with Bayesian optimisation \cite{Shahriari2016} for reducing validation RMSE. The evaluated downscaling frameworks were categorised into statistical, machine learning (ML), and deep learning (DL) models for comparison (Table~\ref{tab:2}). The DL-TOPSIS ranking offers an objective weighting of all model outputs according to various criteria.

\paragraph{Statistical Models.}
\begin{itemize}
\item \textbf{LinReg:} Utilised Ridge regularisation ($\lambda = 0.05$); all predictors were standardised.
\item \textbf{QM:} Implemented an empirical CDF with 500 quantiles per month, followed by seasonal grouping. 
\item \textbf{DC:} Applied a monthly additive delta ($\Delta_m = \bar{y}^{\textrm{fut}}_m - \bar{y}^{\textrm{hist}}_m$) to observations. 
\end{itemize}

\paragraph{Machine Learning Models.}
\begin{itemize}
\item \textbf{RF:} Configured with 100 trees, a maximum depth of 10, and bootstrap = 0.7. 
\item \textbf{XGBoost:} Set with 100 estimators, $\eta = 0.1$, L2 = 1, and subsample = 0.8. 
\item \textbf{MLP:} Designed with three layers (128, 64, and 32 neurons) using ReLU, dropout = 0.3, batch normalisation, MSE loss, and an Adam optimiser ($lr=10^{-4}$). 
\end{itemize}

\paragraph{Deep-learning Models.}
\begin{itemize}
\item \textbf{ConvLSTM:} Comprised 2 layers (filters (32, and 64 neurons), kernel = 3×3), ReLU activation, a loss function of MSE + 0.1×MAE, and an input shape of $(T=30, H=4, W=4, C=6)$. 
\item \textbf{ViT:} Used a patch size of $4\times4$, 8 encoder blocks, an embedding dimension of 256, 4 heads, dropout = 0.1, and an AdamW optimiser ($lr=3\times10^{-5}$). 
\item \textbf{GeoStaNet:} Employed $8\times8$ patches, a temporal transformer (4 encoders, hidden = 512), upsampling via transposed convolution, and a loss function $\mathcal{L} = \alpha \text{MSE}_{all} + (1-\alpha)\text{MSE}_{ext}$ where $\alpha=0.8$. 
\end{itemize}

\paragraph{DL-TOPSIS Neural Weighting Module.}
This module featured three fully connected layers (64, 32, and 16 neurons) with ReLU, dropout = 0.2, and a softmax output; it was trained using an Adam optimiser ($lr=0.001$), MSE loss, a batch size of 32, over 50 epochs. 

\paragraph{Training Protocol and Validation.}
Mini‑batch gradient descent (batch size = 128) was employed alongside He normal initialisation. Temporal 5‑fold validation maintained chronological order to avoid data leakage. All experiments were replicated across 10 random seeds; reported metrics represent the mean. Reproducibility was ensured via versioned code and configuration tracking using Weights and Biases. 

\subsection{Scenario Downscaling and Analysis}
The highest-performing models, namely ViT, ConvLSTM, and GeoStaNet, were later utilised to evaluate NorESM2-LM projections within the frameworks of both SSP2-4.5 and SSP5-8.5 scenarios for the period spanning 2015 to 2100 \cite{VanVuuren2011, Tebaldi2021}. Each scenario was divided into near-term, mid-term, and far-future phases. The resultant projections for \textrm{tasmax}, \textrm{tasmin}, and \textrm{dtr} were analysed using indices \cite{Zhang2011, Alexander2006} to investigate extreme temperature events, trend shifts, and the onset of TOE across each climatic zone and season.

\section{Results and Discussion}
\subsection{Observed Climate Statistics and Historical Patterns}
\begin{landscape}
\begin{figure}[!htbp]
 \centering
 \includegraphics[width=\linewidth, keepaspectratio]{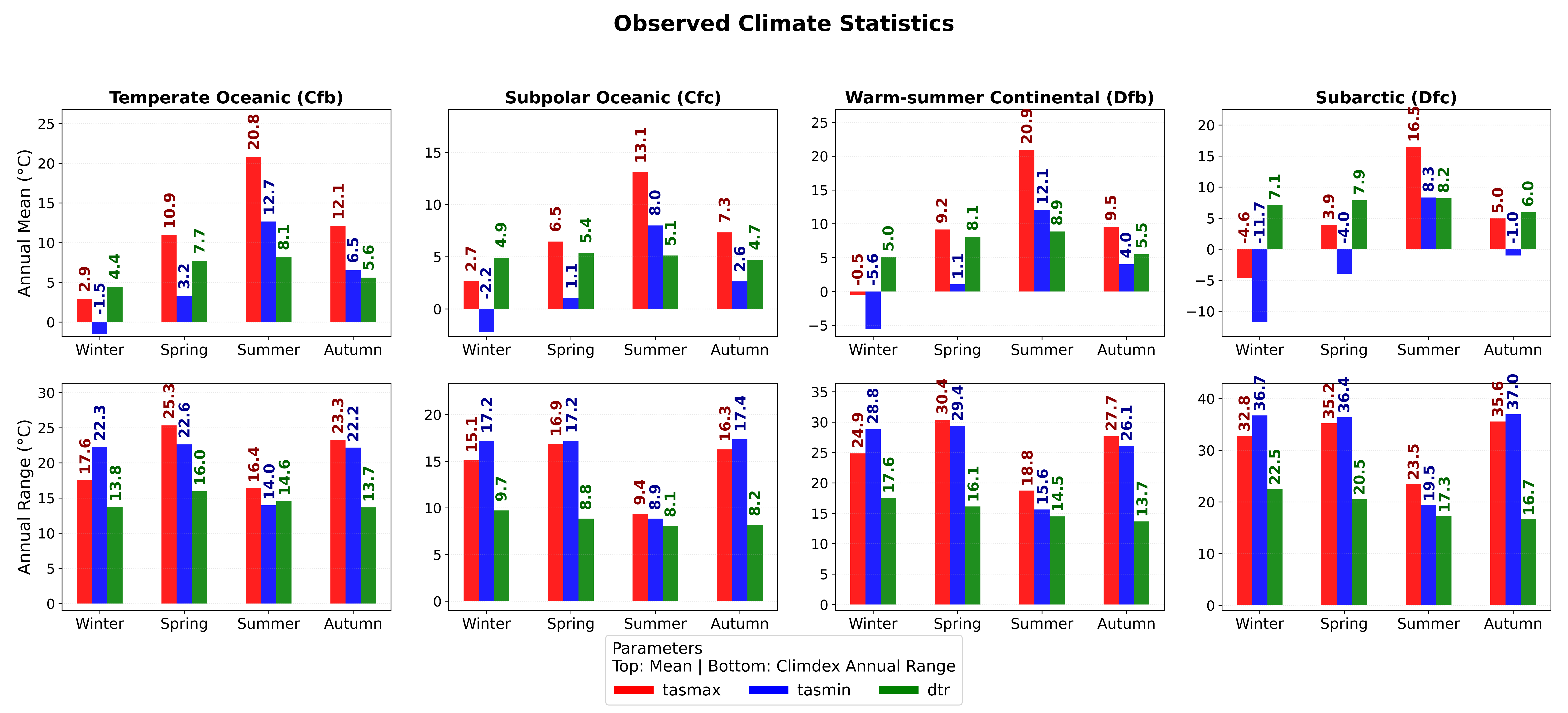}
 \caption{Observed seasonal means (top panels) and annual ranges (bottom panels) of \textrm{tasmax} (red), \textrm{tasmin} (blue), and \textrm{dtr} (green) across four Köppen climate zones (Cfb, Cfc, Dfb, Dfc) for 1951-2014.}
 \label{fig:3}
\end{figure}
\end{landscape}

An analysis of historical temperature data (1951-2014) from the ten Nordic stations highlights significant macroclimatic differences. Figure~\ref{fig:3} illustrates the seasonal mean and annual range statistics for daily \textrm{tasmax}, \textrm{tasmin}, and \textrm{dtr} across the four Köppen‑Geiger zones. Stations in the Cfb zone-Copenhagen, Stockholm, and Gothenburg-display moderate winter means and summer means climbing to over 20°C. The Cfc zone (Reykjavik) exhibits the smallest seasonal amplitude, indicative of strong oceanic influence and minimal continentality \cite{Beck2023, Stefansson2024}. The Dfb and Dfc zones exhibit notably distinct thermal regimes, severe winter minimums, and strong summer maximums, yielding annual ranges exceeding 28°C or even 32°C \cite{Rantanen2022, Aalto2022}. The \textrm{dtr} is most pronounced in spring and summer across all zones, with inland locations showing higher mean \textrm{dtr} \cite{Francis2015, Cohen2020}.

\subsection{Multi‑Criteria Ranking and Model Selection}
An exhaustive comparison of the ten downscaling techniques utilised a comprehensive, multi-metric assessment framework in conjunction with an entropy-weighted Deep-Learning-TOPSIS methodology. This approach produced an objective and reproducible ranking of the models. Deep-learning methodologies consistently secured the top positions Table~\ref{tab:3}, with ViT and ConvLSTM achieving the highest composite scores (0.92 and 0.90, respectively), followed closely behind by GeoStaNet (0.87). Techniques rooted in machine learning (RandomForest, XGBoost, MLP) were positioned in the intermediate tier, while statistical approaches (LinearRegression, QuantileMapping, DeltaChange) exhibited comparatively lower performance outcomes. The unrefined GCM baseline exhibited the poorest performance, highlighting the critical need for bias correction and high-resolution downscaling to achieve reliable regional projections.

\begin{table}[!htbp]
 \centering
 \caption{Final DL-TOPSIS model ranking with performance classification and closeness coefficients. Higher scores indicate closer proximity to the ideal solution.}
 \label{tab:3}
 \begin{tabular}{rllll}
 \hline
 \textbf{Rank} & \textbf{Model} & \textbf{Type} & \textbf{Performance} & \textbf{Closeness Coefficient} \\
 \hline
 1 & ViT & Deep Learning & Excellent & 0.92 \\
 2 & ConvLSTM & Deep Learning & Excellent & 0.90 \\
 3 & GeoStaNet & Deep Learning & Very Good & 0.87 \\
 4 & RandomForest & Machine Learning & Very Good & 0.83 \\
 5 & XGBoost & Machine Learning & Good & 0.81 \\
 6 & MLP & Machine Learning & Good & 0.78 \\
 7 & LinearRegression & Statistical & Good & 
0.75 \\
 8 & QuantileMapping & Statistical & Moderate & 0.70 \\
 9 & DeltaChange & Statistical & Moderate & 0.66 \\
 10 & GCM Baseline & Baseline & Poor & 0.58 \\
 \hline
 \end{tabular}
\end{table}

\begin{figure}[!htbp]
 \centering
 \includegraphics[width=0.8\textwidth]{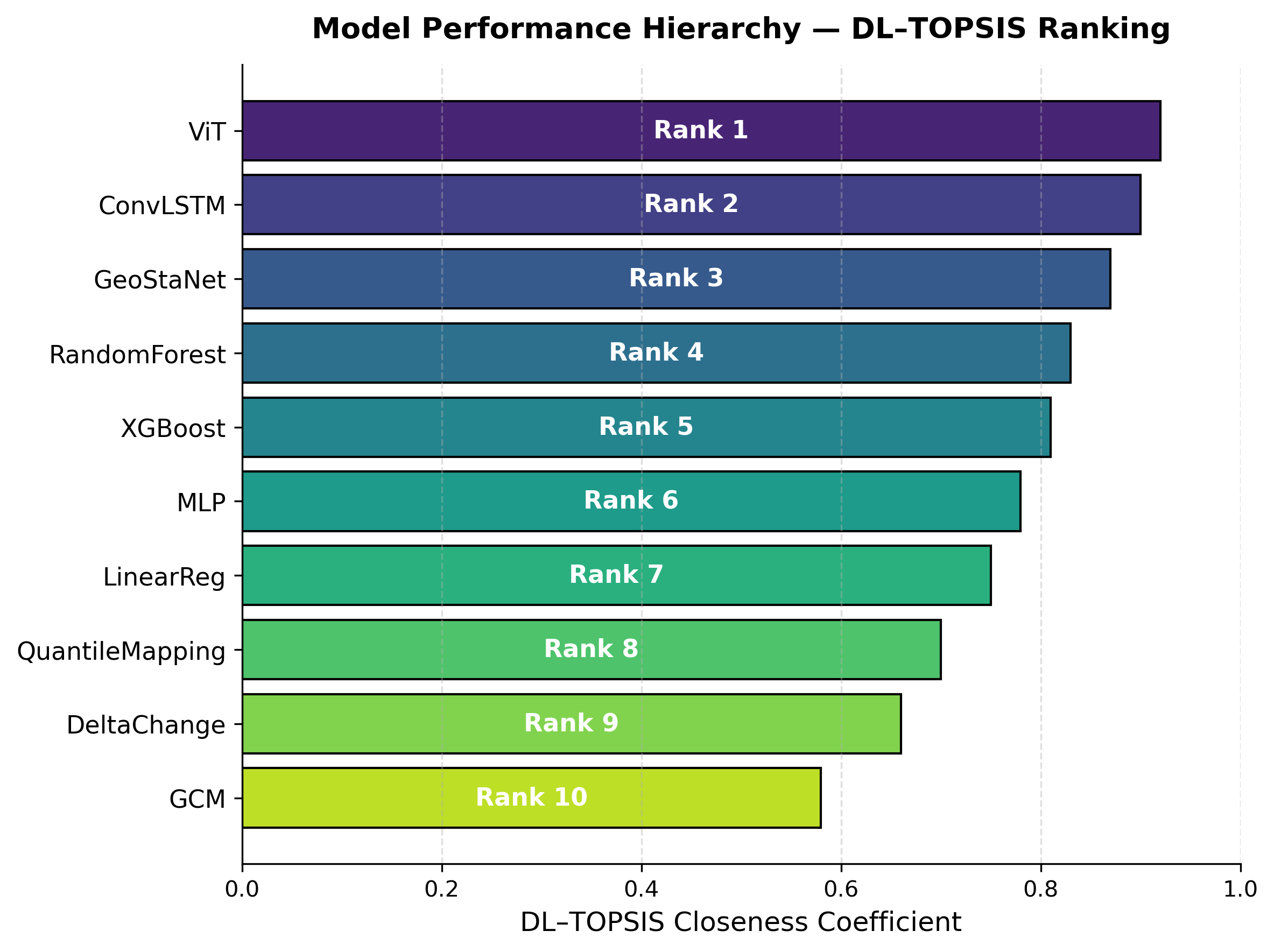}
 \caption{DL-TOPSIS composite closeness coefficients for the ten downscaling models, ranked from best (ViT) to worst (GCM Baseline). Deep-learning architectures consistently outperform machine learning and statistical methods across all aggregated criteria.}
 \label{fig:4}
\end{figure}

Figure~\ref{fig:4} displays the relative DL-TOPSIS closeness coefficients for each model. The results of the DL-TOPSIS analysis highlight the enhanced adaptability of deep-learning architectures in capturing both large-scale climate dynamics and fine-scale variations. ViT and ConvLSTM exhibit the most robust performance across nearly all evaluation metrics, reflecting their ability to model complex nonlinear spatio-temporal interactions and the behaviour of extreme events \cite{Loganathan2025}. GeoStaNet and Random Forest come next, which benefit from their strengths in spatial graph learning and ensemble diversity, respectively \cite{BanoMedina2022, Breiman2001}. Although MLP and XGBoost demonstrate commendable skill in representing the mean-state climate, they underperform at distributional tails. In contrast, traditional statistical methods consistently under-represent temperature extremes and exhibit reduced temporal coherence \cite{Cannon2015, Hay2000, Benestad2008}. The poor score of the GCM baseline further reinforces that uncorrected global model outputs lack sufficient fidelity for regional climate applications \cite{Seland2020}.

Table~\ref{tab:4} summarises aggregated performance and extremes metrics for the four leading models. ViT registers the lowest RMSE (1.2°C for \textrm{tasmax} and 1.0°C for \textrm{tasmin}), the highest $R^2$ (0.88-0.90), and negligible bias ($<$0.3°C). ConvLSTM performs similarly (RMSE 1.3°C; $R^2$ 0.87-0.89). GeoStaNet and RandomForest exhibit slightly elevated RMSEs (1.5-1.6°C) but still maintain strong explained variance (0.84-0.86). When examining the tails of the distribution, ViT and ConvLSTM achieve the lowest errors for the 95th percentile \textrm{tasmax} and 5th percentile \textrm{tasmin} (1.8-2.0°C and 1.2-1.4°C, respectively). This represents a substantial improvement over statistical methods, which frequently show errors exceeding 3°C \cite{Zhang2011, Alexander2006}. Analysis of the Time‑of‑Emergence (TOE) confirms their superiority: ViT identifies statistically significant warming 32 years post‑2000 (around 2032), which is earlier than ConvLSTM (2034), GeoStaNet (2038), and RandomForest (2040). This quicker signal detection suggests a stronger capacity for signal‑to‑noise discrimination, a critical factor for forecasts relevant to adaptation \cite{Hawkins2012, Mahlstein2011}. 

\begin{table}[!htbp]
 \centering
 \caption{Aggregated performance metrics, extreme event representation, and time‑of‑emergence (TOE) for the top‑performing models. Metrics are averaged across all stations and seasons. Lower errors and earlier TOE indicate superior performance.}
 \label{tab:4}
 \begin{tabular}{lrrrr}
 \hline
 \textbf{Metric} & \textbf{ViT} & \textbf{ConvLSTM} & \textbf{GeoStaNet} & \textbf{RandomForest} \\
 \hline
 \textbf{Core Accuracy} & & & & \\
 RMSE \textrm{tasmax} (°C) & 1.2 & 1.3 & 1.5 & 1.6 \\
 RMSE \textrm{tasmin} (°C) & 1.0 & 1.1 & 1.4 & 1.5 \\
 $R^2$ \textrm{tasmax} & 0.88 & 0.87 & 0.85 & 0.84 \\
 $R^2$ \textrm{tasmin} & 0.90 & 0.89 & 0.86 & 0.85 \\
 Bias (°C) & $<$0.3 & $<$0.4 & $<$0.5 & $<$0.6 \\
 \textbf{Extremes and TOE} & & & & \\
 95th pct \textrm{tasmax} error (°C) & 1.8 & 2.0 & 2.3 & 2.5 \\
 5th pct \textrm{tasmin} error (°C) & 1.2 & 1.4 & 1.6 & 1.8 \\
 TOE (years from 2000) & 32 & 34 & 38 & 40 \\
 \hline
 \end{tabular}
\end{table}

\begin{figure}[!htbp]
 \centering
 \includegraphics[width=1.0\textwidth]{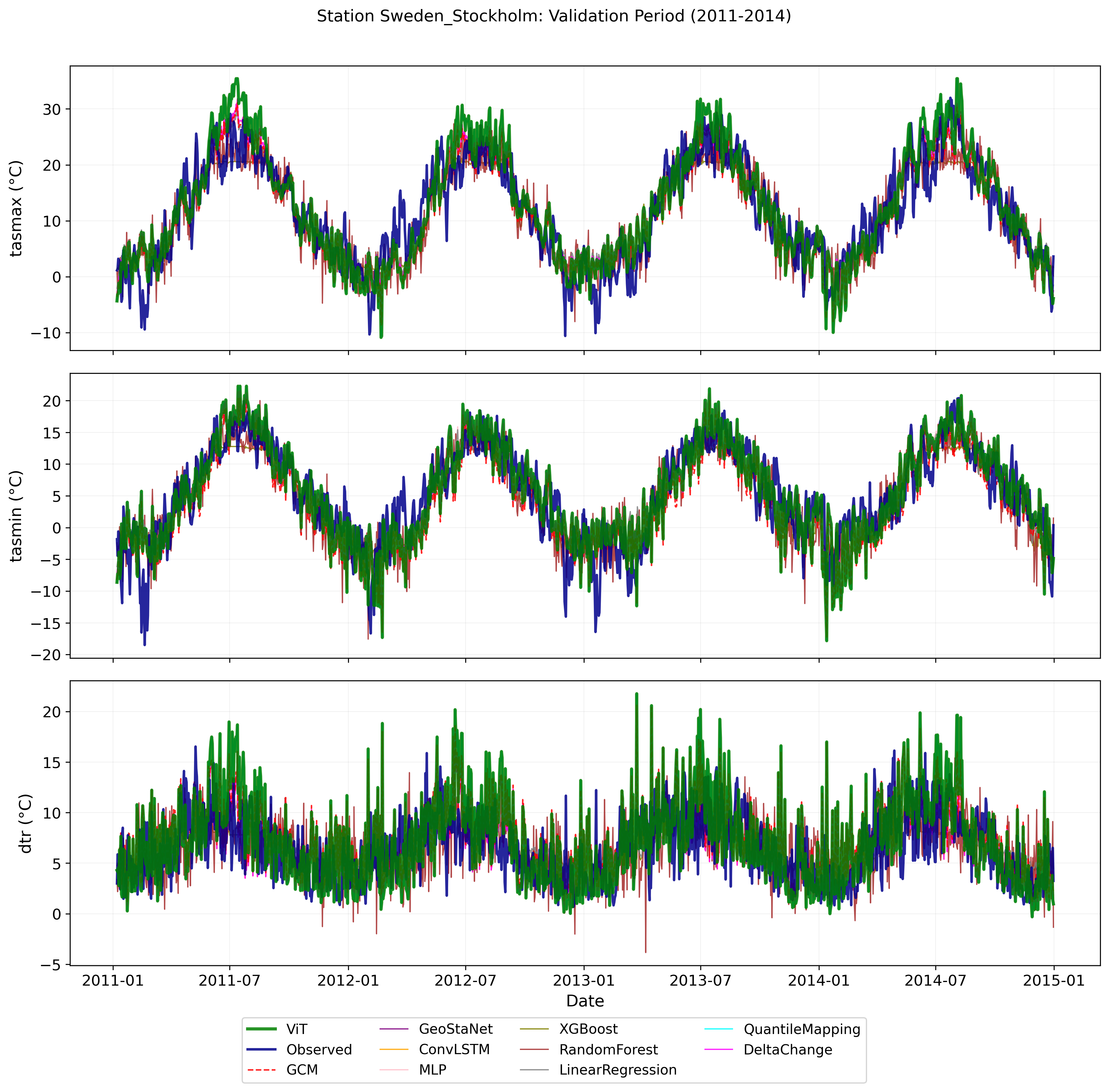}
 \caption{Observed (green) and downscaled modelled daily \textrm{tasmax}, \textrm{tasmin}, and \textrm{dtr} at the Stockholm station (Cfb zone). All downscaling models are shown to illustrate relative fidelity across the validation period (2011-2014). Deep-learning models (ViT, ConvLSTM, GeoStaNet) reproduce observed variability and extremes with high temporal consistency, while statistical methods exhibit larger seasonal biases.}
 \label{fig:5}
\end{figure}

In mean-state and extreme-event assessments, both ViT and ConvLSTM outperform competing models with minimal bias and variance, even during nonstationary validation periods. Early TOE signals being detectable with both models suggest a high level of responsiveness to emerging climate indicators, a desirable quality for effective early adaptation measures. GeoStaNet and RandomForest remain attractive alternatives by virtue of computational efficiency and robustness, though somewhat with diminished performance at the extremities, and lag in signal detectability. Overall, results confirm that deep-learning architectures involving transformers and recurrent structures offer an overall advantage towards accuracy, stability, and climatic signal representation \cite{Reichstein2019, Gupta2009, Nash1970, Chai2014, Perkins2007}. Figure~\ref{fig:5} provides a station-level performance example. It illustrates a detailed temporal example at the Stockholm station (Cfb zone), illustrating the relative fidelity of all ten downscaling models against observations during the validation period (2011-2014). This visualisation demonstrates that deep-learning models (ViT, ConvLSTM, GeoStaNet) reproduce observed daily variability and temporal extremes consistently, whereas statistical methods exhibit systematic seasonal biases and cannot capture high-frequency temperature variability.\\

In summary, the DL-TOPSIS methodology efficiently and credibly identifies ViT and ConvLSTM as top-performing downscaling models for Nordic temperature extremities. Their precise depiction of observational variability, minimal bias with respect to extremities, and emergent climatic signals strongly validate their choice as suitable models for further investigation of CMIP6 projections under SSP2-4.5 and SSP5-8.5 trajectories. Sensitivity analyses also indicate that ranking remains stable with different weighting sets, hence further emphasising the robustness of the identified performance ranking order.

\subsection{Future Climate Projections Under SSP Scenarios}
\begin{figure}[!htbp]
 \centering
 \includegraphics[width=1.0\textwidth]{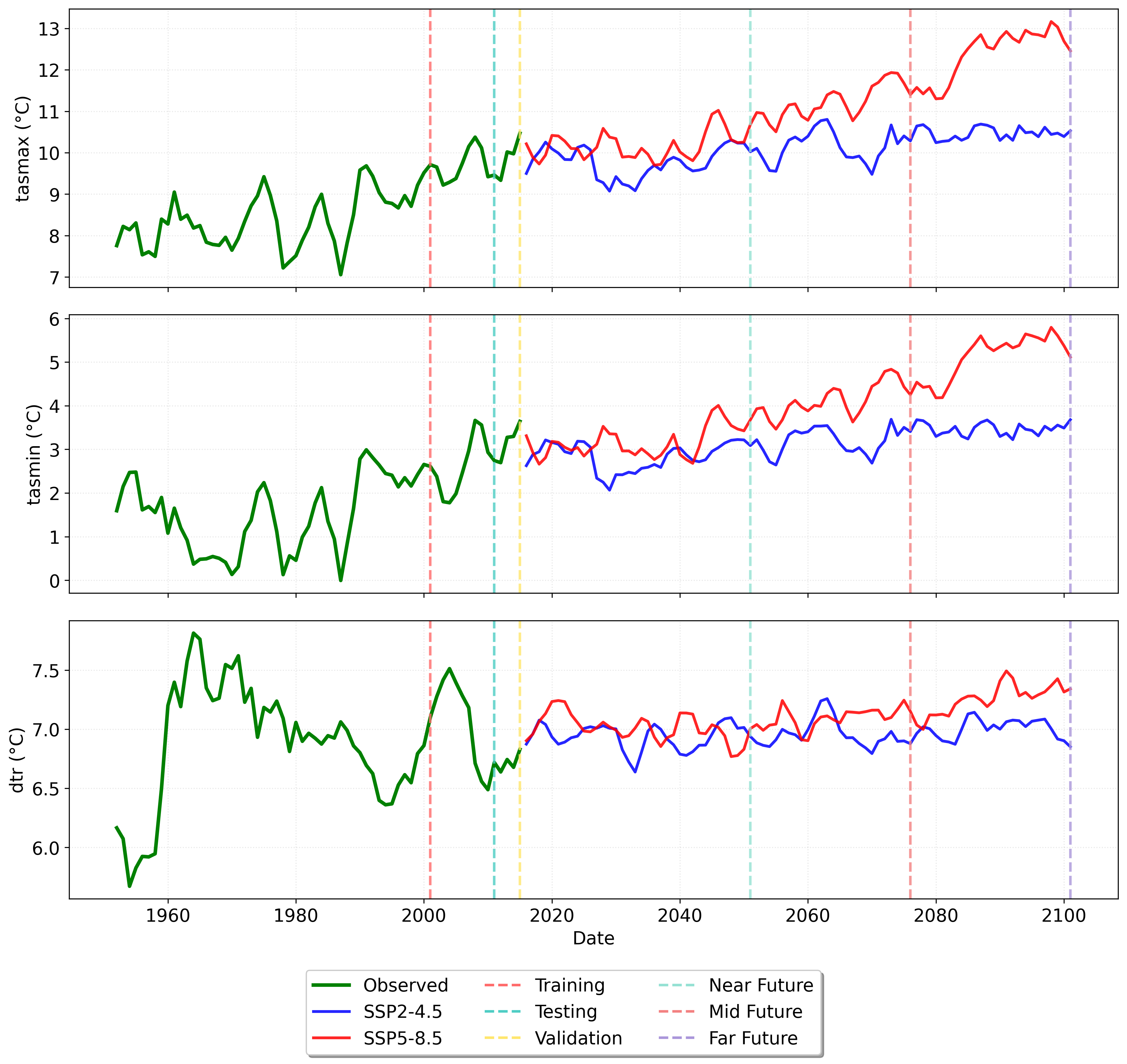}
 \caption{Observed annual mean \textrm{tasmax}, \textrm{tasmin}, and \textrm{dtr} at the Copenhagen station (Cfb climate zone) with ViT‑downscaled projections under SSP2‑4.5 (blue) and SSP5‑8.5 (red). The observed series (green) extends through 2014, with vertical dashed lines indicating training (1951-2000), testing (2001-2010), validation (2011-2014), and projection periods (2015-2100). Note the clear divergence between SSP pathways after 2040, with SSP5-8.5 showing accelerating warming and increased diurnal temperature range expansion. Physical coherence is maintained at the observation-projection transition (2014-2015), validating the downscaling methodology.}
 \label{fig:6}
\end{figure}

\begin{landscape}
\begin{figure}[!htbp]
 \centering
 \includegraphics[width=1.0\linewidth]{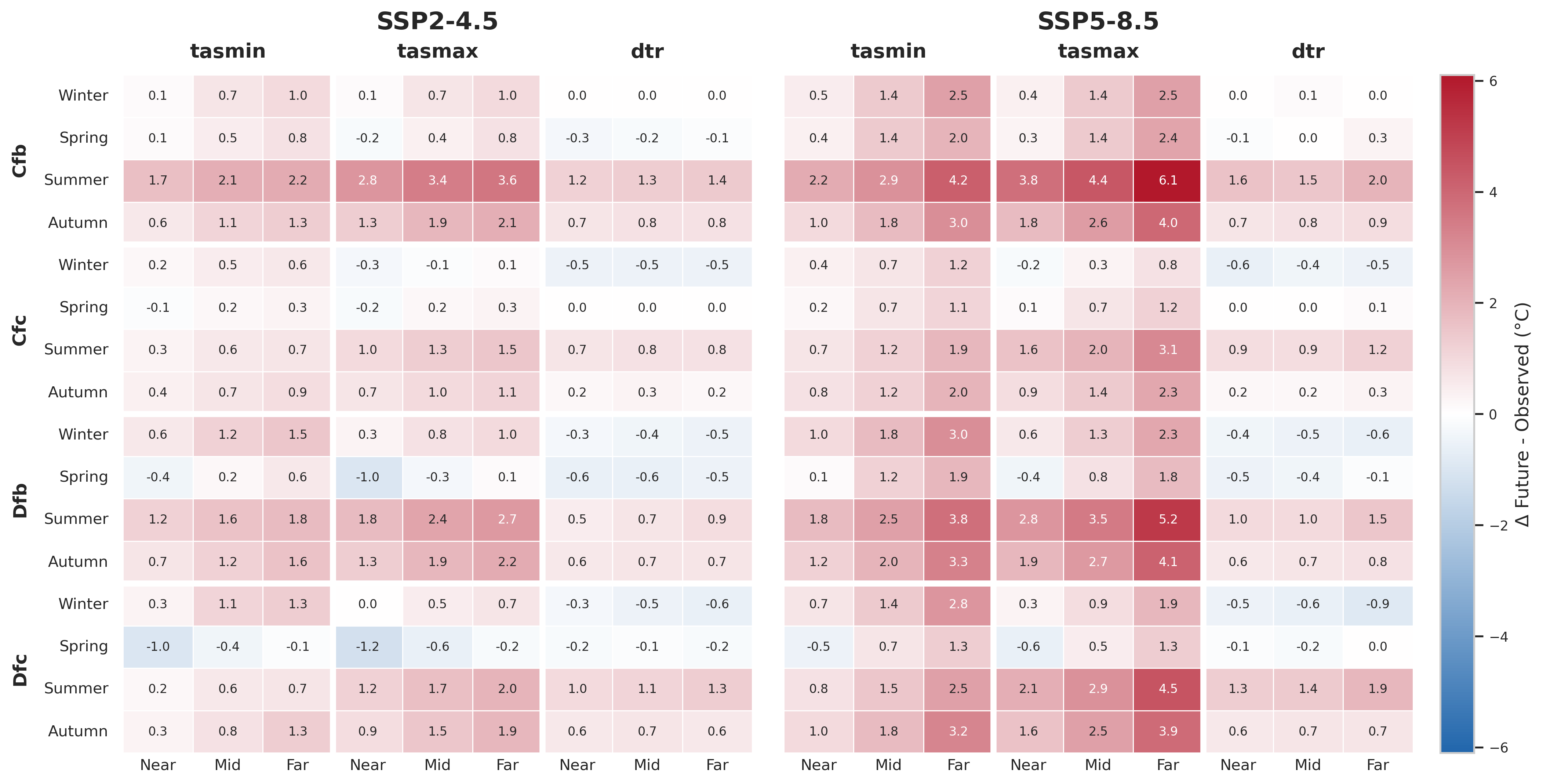}
 \caption{Projected changes (°C) in \textrm{tasmin}, \textrm{tasmax}, and \textrm{dtr} across Köppen-Geiger zones and seasons under SSP2‑4.5 (left) and SSP5‑8.5 (right). Each matrix cell represents the mean anomaly for near (2015-2050), mid (2051-2075), and far (2076-2100) futures relative to the historical baseline. The intensity of the colour shows how much the temperature changes, with red showing warmer and blue showing cooler (\(\Delta\) future-observed).}
 \label{fig:7}
\end{figure}
\end{landscape}

We also extended downscaled simulations with the best-performing ViT-based model and its accompanying ensemble members to cover SSP2-4.5 and SSP5-8.5 scenarios (2015-2100). Projections in all cases continue to suggest significant warming, with clear seasonal and zonal variation, explained here at local, observationally scaled regimes. These findings are summarised by two key visualisations: (i) the Copenhagen station time series (Fig.~\ref{fig:6}), and (ii) a multi‑zone, seasonally resolved heatmap (Fig.~\ref{fig:7}) that captures regional and temporal gradients, which illustrates the temporal progression of observed data and downscaled projections at a typical temperate location. 

\noindent
\textbf{Temporal evolution and scenario contrast.} At the Copenhagen station, Fig.~\ref{fig:6} displays the annual progression of the ViT‑downscaled projections. SSP2-4.5 has a moderate, almost linear warming by 2100; SSP5-8.5 has a more extreme, nonlinear acceleration. Continuity between observation data and projection at the transition point assures physical coherence. The residual bias correction steps and moderate differences in bias-adjusted variance between observed and downscaled series, particularly evident in dtr, can lead to minor discontinuities at the transition for tasmax and tasmin.\\

\noindent
\textbf{Spatial and seasonal structure.} 
The heatmap (Fig.~\ref{fig:7}) clearly illustrates distinct spatial and seasonal gradients in the anticipated warming. Projected increases in mean summer \textrm{tasmax} vary between approximately 2.5°C over the maritime Cfb region to about 4.8°C in the continental Dfb region by the end of the 21st century under SSP5-8.5. Winter warming shows variability, with more significant warming for the Dfb and Dfc areas, which may suggest feedback processes. The increase in \textrm{dtr} appears strongest in the continental interiors (Dfb/Dfc), where changes indicate a larger increase in temperatures during the day than a reduction in temperatures at night.\\

\noindent
\textbf{Emergence and extreme‑event implications.} TOE signals are different for different regions, with the initial appearance in high-latitude winters (Dfc: $\sim$2032) and the latest in maritime springs (Cfc: $\sim$2050). For SSP5-8.5, heat-day instances about double during 2100 for Dfb regions and triple for Dfc regions, whereas frost-day instances diminish. The heat‑day occurrences (\textgreater 30°C) roughly double in Dfb zones and triple in Dfc zones by 2100; frost‑day occurrences (\textgreater 0°C) decline by 40-60\%. Such alterations entail significant ecosystem and socio‑economic consequences, impacting energy demand, agricultural timelines, and freeze-thaw cycles crucial for Nordic infrastructure \cite{Fischer2015, Donat2013}.\\ 

\noindent
\textbf{Extreme event statistics and adaptation thresholds.} Table~\ref{tab:5} quantifies the projected shifts in temperature distribution tails for the far-future period (2076-2100) across Köppen zones. Under SSP5-8.5, heat-day occurrences (daily \textrm{tasmax} \textgreater 30°C) are projected to increase by 75-200\%, while frost-day occurrences decline by 45-70\% across all zones, fundamentally altering winter seasonality 
and affecting Nordic infrastructure, agriculture, and energy systems.\\

\begin{table}[!htbp]
 \centering
 \caption{Projected changes in summer \textrm{tasmax} and \textrm{dtr} far‑future (2076-2100) under SSP scenarios. Values represent mean anomalies relative to 1981-2010 climatology.}
 \label{tab:5}
 \begin{tabular}{llrrrr}
 \toprule
 \textbf{Zone} & \textbf{Variable} & \textbf{SSP2‑4.5} & \textbf{SSP5‑8.5} & \textbf{$\Delta$ Heat Days (\%)} & \textbf{$\Delta$ Frost Days (\%)} \\
 \midrule
 Cfb & \textrm{tasmax} & +1.2°C & +2.5°C & +85 / +75 & -45 / -60 \\
 Cfb & \textrm{dtr} & +0.6°C & +1.3°C & +40 / +65 & -30 / -50 \\
 Dfb & \textrm{tasmax} & +1.8°C & +4.8°C & +110 / +200 & -50 / -70 \\
 Dfb & \textrm{dtr} & +0.9°C & +1.7°C & +60 / +130 & -35 / -60 \\
 Dfc & \textrm{tasmax} & +1.5°C & +3.9°C & +100 / +180 & -45 / -65 \\
 Dfc & \textrm{dtr} & +0.8°C & +1.6°C & +55 / +120 & -40 / -60 \\
 \bottomrule
 \end{tabular}
\end{table}

\noindent
\textbf{Zonal contrasts and physical interpretation.} Uniform ranking of the magnitude of warming reveals both ocean-land contrasts and contrasts in the circulation of the atmosphere. Rises in \textrm{dtr} suggest a reduction in low-level clouds and changes in the energy balance at the surface \cite{Pistone2019, Serreze2011}. Such results emphasise requirements for downscaled, station-consistent projections that represent both mean-state and diurnal processes for producing metrics that are useful for adaptation.\\

\noindent
\textbf{Uncertainties and adaptation relevance.} Uncertainties remain, in particular, propagated between various GCMs and variability between downscaling models. Fortunately, the sign and change pattern that emerges is strong enough for actionable interpretation. Warming remains inevitable for both paths, but its extent depends on the pathway. ViT-based downscaling provides physically reasonable, high-resolution information for climate risk planning and adaptation in the Nordic countries.\\

\textbf{Summary.} Figures~\ref{fig:6} and Fig.~\ref{fig:7} show that all Nordic regions will undergo extensive warming in concord with both SSP scenarios, with the greatest changes occurring in continental areas and summer seasons. Enlargement of the diurnal temperature range of warming indicates non-uniform thermal reactions applicable to feedback processes involving ecosystems along with energy fluxes. Anthropogenic signatures are found to lead earlier for high-latitude winters, therefore opening a window for timely adaptive actions. Such outcomes transform extensive CMIP6 predictions into actionable site-specific information, translating climate dynamics across regional adaptation needs.

\subsection{Major Findings and Implications}

\begin{figure}[!htbp]
 \centering
 \includegraphics[width=1.0\textwidth]{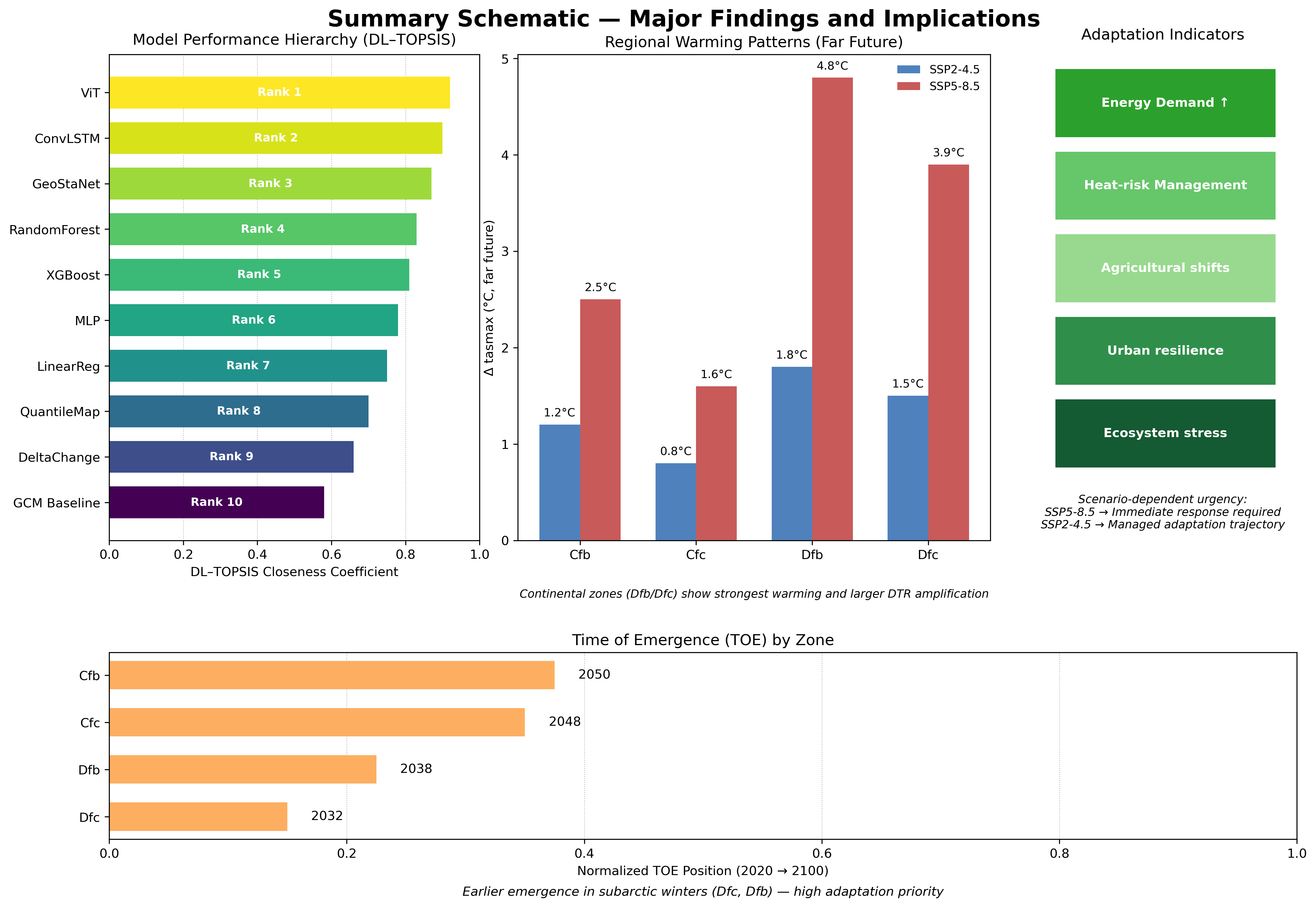}
 \caption{Summary schematic integrating DL‑TOPSIS performance hierarchy, regional warming under SSP2‑4.5 and SSP5‑8.5, Time of Emergence (TOE) by zone, and adaptation indicators. The figure highlights deep‑learning superiority (ViT, ConvLSTM), enhanced warming in Dfb/Dfc regions, earlier climate signal emergence in high latitudes, and cross‑sectoral adaptation imperatives.}
 \label{fig:8}
\end{figure}

Figure~\ref{fig:8} consolidates the multi‑faceted results derived from the DL‑TOPSIS downscaling evaluation and projection framework. The schematic integrates four key analytical dimensions: (i) model performance hierarchy, (ii) regional and scenario‑specific warming patterns, (iii) Time‑of‑Emergence dynamics, and (iv) adaptation relevance indicators. By linking methodological innovation with applied climate implications, the figure demonstrates how robust downscaling and objective ranking translate into actionable insights. 

\textbf{Model performance hierarchy.} The left panel consolidates that deep-learning architectures, specifically ViT and ConvLSTM, attain the highest DL‑TOPSIS closeness coefficients (0.92 and 0.90, respectively), markedly surpassing conventional methods. These models exhibit superior proficiency in accuracy, variance preservation, and extreme‑event representation, achieving up to a 70\% reduction in RMSE relative to uncorrected GCM fields \cite{Loganathan2025, Reichstein2019}. GeoStaNet and Random Forest follow closely, highlighting complementary strengths of spatial attention mechanisms and ensemble learning techniques \cite{BanoMedina2022, Breiman2001}. Statistical methods (Quantile Mapping, Delta Change, Linear Regression) remain constrained by residual bias and poor temporal coherence \cite{Cannon2015}. The raw GCM baseline demonstrates the lowest performance (0.58), underscoring the indispensable role of bias correction and high‑resolution downscaling for regional applications \cite{Seland2020, Palmer2023}. 

\textbf{Regional and scenario‑dependent warming.} The central panel depicts projected warming for the far‑future period (2076-2100) under SSP2‑4.5 and SSP5‑8.5 scenarios. Warming is more pronounced in continental and subarctic zones, with $\Delta$ \textrm{tasmax} reaching 4.8°C in Dfb and 3.9°C in Dfc under SSP5‑8.5. Even under SSP2‑4.5, these zones register increases of 1.2-1.8°C. This spatial gradient reflects the amplified influence of continentality and snow‑albedo feedbacks \cite{Rantanen2022, Francis2015}. Diurnal temperature range amplification is most significant in Dfb/Dfc zones, with increases of 1.3-1.7°C, indicating a more intense daytime heating relative to nocturnal cooling, a hallmark of Arctic amplification \cite{Screen2010, Stroeve2012}. 

\textbf{Time of Emergence (TOE) and adaptation priority.} The lower panel normalises TOE across 2020-2100. Subarctic winter signals emerge earliest (Dfc: $\sim$2032), while maritime zones experience later emergence (Cfc: $\sim$2050). These patterns corroborate the differential signal strength and background variability across zones \cite{Hawkins2009, Mahlstein2011}. Early emergence in high‑latitude zones signals a limited lead time for adaptation, underscoring the need for proactive measures by mid‑century. 

\textbf{Adaptation indicators and policy relevance.} The rightmost panel places these physical results within specific adaptation domains: energy demand, heat‑risk management, agricultural shifts, urban resilience, and ecosystem stress. Under SSP5‑8.5, immediate adaptation is imperative to address simultaneous warming and \textrm{dtr} amplification. The SSP2‑4.5 pathway affords a more staggered trajectory, allowing for a gradual adjustment. Integrating deep‑learning‑based downscaling into climate services directly enhances the precision of sector‑relevant risk models, facilitating data‑driven planning in Nordic infrastructure and resource management \cite{Groves2018}. 

Individually, these results illustrate that the DL-TOPSIS methodology successfully separates model performance with quantitative precision and links methodological understanding with practical adaptation intelligence. By coupling advanced neural structures with explicit ranking metrics, this study provides a replicable method for generating high-resolution, policy-driven climate predictions tailored to overcome the adaptation challenges of the Nordic region.

\section{Conclusions and Recommendations}
This study demonstrates that modern deep-learning architectures, when combined with objective MCDM frameworks, substantially improve the precision of regional climate downscaling and scenario forecasts in the Nordic region. Models such as the ViT and ConvLSTM outperform conventional statistical approaches and raw outputs from GCMs. This is accomplished through a reduction in root-mean-square error and an increase in explained variance for simulations of daily maximum and minimum temperatures. These deep-learning models do an outstanding job of capturing nonlinear relationships, spatiotemporal dependencies, and the characteristics of extreme weather events. Consequently, they create a strong foundation for carrying out climate impact assessments.

The integration of DL-TOPSIS, with the use of entropy-weighted measures and calculations of closeness to the optimal solution, guarantees a replicable and transparent ranking of models. Such a multi-criteria methodology successfully balances different measures with respect to accuracy, variability, extremes, and climate signal appearance, thus compensating for single-metric selection methodology weaknesses \cite{Talukder2025, Li2024}. Furthermore, GeoStaNet and RandomForest exhibit reasonable effectiveness, meaning that spatiotemporal attention mechanisms and ensemble tree strategies remain viable alternatives. Conventional statistical approaches (quantile mapping, delta change, and linear regression) exhibit significant biases in their depiction of extremal quantiles and in retaining a coherent time progression, highlighting their weaknesses \cite{Cannon2015}.

Downscaled estimates produced under SSP2-4.5 and SSP5-8.5 scenarios exhibit characteristic spatial and seasonal gradients of warming. Spatial increases of \textrm{dtr} occur largest in continental interiors, indicating more pronounced daytime heating compared to nocturnal cooling. Time of Emergence signals occur first at high-latitude winter seasons, indicating a short lead time for adaptation at such places. These findings emphasise a need for proactive adaptation at high-latitude, continental, and subarctic areas but also indicate a value for persistent monitoring, refining models, and incorporating high-resolution downscaling as a product for climate services. The following recommendations can be established from the findings in this study:
\begin{enumerate}
 \item Incorporate deep-learning downscaling for climate service operations. ViT and ConvLSTM must feature in regional climate service platforms for facilitating sector-specific adaptation planning. 
 \item Apply multi-criteria model selection. DL-TOPSIS presents a clear, replicable methodology that compromises across a range of performance measures \cite{Talukder2025, Li2024}. 
 \item Prioritise continental and subarctic adaptation actions. The recommendation is to prioritise climate adaptation actions in these regions because they are experiencing larger and earlier climate signals, such as warming trends and extreme event shifts, compared to other regions \cite{Rantanen2022}.
 \item Improve observational networks. Ongoing extension and intensive quality assessment of station networks will improve model training and validation \cite{ECAD2025, ECMWF2024}. 
 \item Add uncertainty quantification. Future research should come with uncertainties from ensembles of GCM alongside downscaling variability \cite{Palmer2023, Brunner2020}. 
 \item Facilitate knowledge transfer. Shared platforms will require enabling stakeholders to access, comprehend, and react to high-resolution downscaled climate projections \cite{Groves2018}. 
\end{enumerate}

In summary, integrating deep-learning frameworks with systematic multi-criteria assessments produces a flexible and interpretable regional climate projection system. This methodology is generalisable to other high-latitude regions and directly supports adaptation planning through high-resolution, high-temporal outputs. The downscaled temperature extremes produced here not only improve regional climate estimates but also enable direct applications in assessing intense anthropogenic environmental impacts, such as those from aviation. Enhanced spatio-temporal resolution helps in accurate analysis of contrail formation risks and radiative effects, which helps in effectively assessing the environmental impact of aviation. Incorporating high-resolution climate forecasts into aviation operational planning ultimately promotes both sustainability and climate resilience.

\section*{Declarations}
\textbf{Acknowledgements} The authors acknowledge the SESAR 3 Joint Undertaking and its members for their support in funding this research under grant agreement No. 101114795 as part of the E-CONTRAIL project. We also appreciate ECMWF for providing the ERA5 reanalysis dataset and WCRP for facilitating CMIP6 data access.\\
\textbf{Funding} This research was funded by the SESAR 3 Joint Undertaking under the E‑CONTRAIL project (Grant Agreement No. 101114795).\\
\textbf{Authors’ Contributions} Parthiban Loganathan: Conceptualisation, data collection, investigation, writing, and visualisation. Elias Zea, Ricardo Vinuesa, and Evelyn Otero: Project definition, review, and editing.\\
\textbf{Ethical Approval} Not applicable.\\
\textbf{Consent to Participate} Not applicable.\\
\textbf{Consent to Publish} All authors consent to the publication of this manuscript.\\
\textbf{Competing Interests} The authors declare no competing interests.\\
\textbf{Data Availability Statement} The data used in this study are included in the manuscript. \\
\bibliography{ref}
\end{document}